\documentclass[aps,prl,reprint,superscriptaddress]{revtex4-1}

\usepackage{amsmath,amsfonts,amssymb}

\usepackage[pdftex]{graphicx}
\usepackage{gensymb}
\usepackage{microtype}
\usepackage{bm}\let\vec\bm
\usepackage[normalem]{ulem}

\usepackage[svgnames]{xcolor}
\usepackage[
	colorlinks=True,linkcolor=DarkRed,citecolor=ForestGreen,urlcolor=MediumBlue,
	pdfstartview=FitH,bookmarks=False,pdfpagemode=UseNone
]{hyperref}

\newcommand{\NbSe}{NbSe$_3$}

\begin{document}

\title{\boldmath Dimensional Crossover in a Charge Density Wave Material Probed\\ by Angle-Resolved Photoemission Spectroscopy
}

\author{C. W. Nicholson}
\email{nicholson@fhi-berlin.mpg.de}
\affiliation{Department of Physical Chemistry, Fritz-Haber-Institut, Faradayweg 4-6, Berlin 14915, Germany}
\author{C. Berthod}
\affiliation{Department of Quantum Matter Physics, University of Geneva, 24 quai Ernest-Ansermet, 1211 Geneva, Switzerland}
\author{M. Puppin}
\affiliation{Department of Physical Chemistry, Fritz-Haber-Institut, Faradayweg 4-6, Berlin 14915, Germany}
\author{H. Berger}
\affiliation{Institut de la Mati\`ere Complexe,
\'Ecole Polytechnique F\'ed\'erale de Lausanne, 1015 Lausanne, Switzerland}
\author{M. Wolf}
\affiliation{Department of Physical Chemistry, Fritz-Haber-Institut, Faradayweg 4-6, Berlin 14915, Germany}
\author{M. Hoesch}
\affiliation{Diamond Light Source, Harwell Campus, Didcot OX11 0DE, Oxfordshire, United Kingdom}
\author{C. Monney}
\affiliation{Institute of Physics, University of Zurich, Winterthurerstrasse 190, 8057 Zurich, Switzerland}

\date{May 18, 2017}

\begin{abstract} 
High-resolution angle-resolved photoemission spectroscopy (ARPES) data reveal evidence of a crossover from one-dimensional (1D) to three-dimensional (3D) behavior in the prototypical charge density wave (CDW) material \NbSe{}. In the low-temperature 3D regime, gaps in the electronic structure are observed due to two incommensurate CDWs, in agreement with x-ray diffraction and electronic-structure calculations. At higher temperatures we observe a spectral weight depletion that approaches the power-law behavior expected in 1D. From the warping of the quasi-1D Fermi surface at low temperatures, we extract the energy scale of the dimensional crossover. This is corroborated by a detailed analysis of the density of states, which reveals a change in dimensional behavior dependent on binding energy. Our results offer an important insight into the dimensionality of excitations in quasi-1D materials.

\end{abstract}

\maketitle

In one spatial dimension (1D), reduced screening and a restricted phase space for scattering heavily impact the electronic properties of materials due to the ensuing strong correlations. As a result, the well-known concept of Fermi liquid (FL) breaks down, and may be replaced by the Tomonaga-Luttinger liquid (TLL) \cite{Tomonaga1950, Luttinger1963}, in which correlation functions display power-law behavior. The fundamental excitations of a TLL are collective bosonic modes carrying only spin or charge, rather than electron-like fermionic quasi-particles \cite{Voit1993, Voit1994, Giamarchi2003}. Furthermore, long-range ordered phases are not stable in a purely 1D system as a result of quantum and thermal fluctuations \cite{Mermin1966}; thus a dimensional crossover should be a prerequisite for a 1D system to enter an ordered phase, as occurs in a number of quasi-1D materials \cite{Gruner1988, Gruner1994}. This is distinguished by a crossover energy, $E_{\mathrm{C}}$, or temperature above which excitations exhibit 1D character, while low-energy excitations behave as in a FL \cite{Castellani1994, Arrigoni1999, Biermann2001}. In photoelectron spectroscopy experiments, spectral weight depletion near the Fermi energy has been interpreted as a characteristic of TLL behavior in a variety of systems \cite{Dardel1991,Gweon2004}, also at very low temperatures \cite{Ishii2003,Ohtsubo2015}. In contrast, in quasi-1D systems, power-law correlations are expected to be observed only above the dimensional crossover energy or temperature. To date, the properties of the low-temperature phase, in particular how strong one-dimensional correlations are imprinted on it, remain poorly understood \cite{Giamarchi2004}.

In this Letter, we report a high-resolution ARPES study of \NbSe{} single crystals, including the evolution of the electronic structure over a wide temperature range and a mapping of the Fermi surface. We find evidence of a dimensional crossover from 1D to 3D as a function of decreasing energy and temperature. CDW gaps in the electronic structure are observed at low temperatures, occurring at momenta consistent with x-ray data and reproduced by theoretical simulations. Conversely at high temperatures a power-law suppression of the spectral function is observed, suggestive of 1D behavior. From the warping of the Fermi surface measured at low temperature, a crossover energy scale of around $E_{\mathrm{C}} \approx 110$~meV (1250~K) is extracted based on a tight-binding model. This is corroborated by an analysis of the density of states which reveals 1D behavior only above $E_{\mathrm{C}}$. The data presented here attest an intermediate regime in which the bosonisation expected for a purely 1D dispersion is still partially observed while approaching the FL regime of 3D coherence.

\NbSe{} is an archetypical linear-chain compound, which undergoes CDW transitions at $T_1=145$~K and $T_2=59$~K \cite{Monceau1977} with incommensurate modulation wave vectors $\vec{q}_1=(0,0.243,0)$ and $\vec{q}_2=(0.5,0.263,0.5)$ respectively, in units of the reciprocal lattice parameters ($a^{*},b^{*},c^{*}$) \cite{Fleming1978, Hodeau1978}. The occurrence of the CDW has been ascribed to Fermi-surface nesting \cite{Schafer2001,Schafer2003a}. Despite intense research on \NbSe{} (for an overview see Refs.~\cite{Gruner1988, Monceau2012}), detailed information about the electronic dispersion is limited to only a few studies \cite{Schafer2001,Schafer2003a} by ARPES. Recent work by scanning tunneling microscopy showed a surface CDW transition temperature higher than that in bulk, and confirmed the higher dimensional nature of this material at low temperatures \cite{Brun2009, *Brun2010} which had previously been observed in x-ray scattering data \cite{Moudden1990}. In contrast, intriguing reduced dimensional behavior at the surface has also been revealed as signatures of soliton behavior \cite{Brazovskii2012}.

Single crystals of \NbSe{} of typical dimensions $20\times500~\mu$m$^{2}$ were cleaved in vacuum at a pressure lower than $5\times10^{-11}$~mbar. ARPES measurements were carried out at the IO5 beam line of the Diamond Light Source \cite{Hoesch2017} over a temperature range 6.5--260~K with photon energies 20--40~eV (linear horizontal polarization). The angular and energy resolution were $0.2\degree$ and 10~meV, respectively. All results presented here have been reproduced by measurements on multiple samples.

\begin{figure}[tb]
\includegraphics[width=\columnwidth]{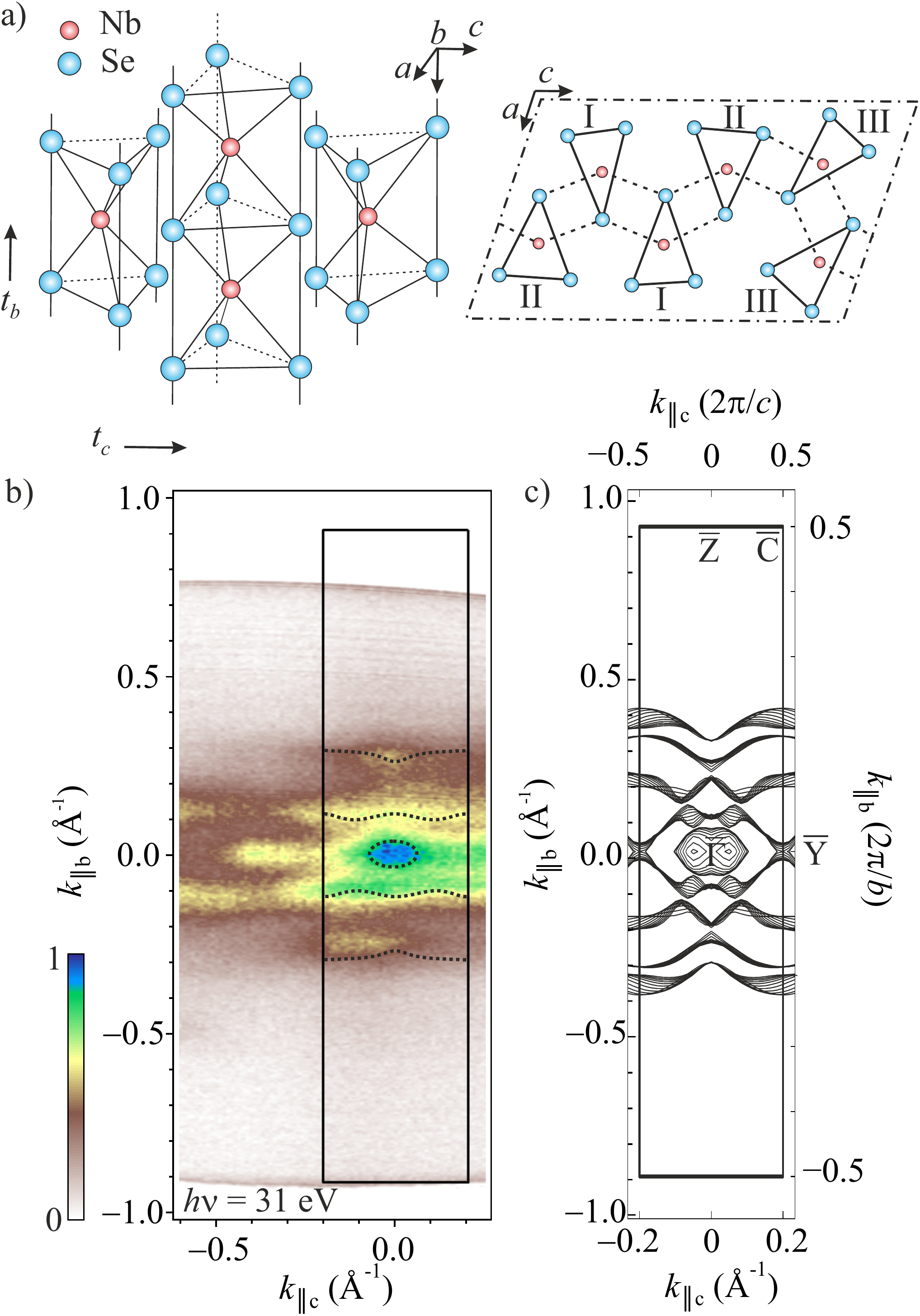}
\caption{\label{fig:Figure 1}
(Color online) (a) Schematic crystal structure of \NbSe{} (adapted from \cite{Monceau1977}). $t_{b}$ and $t_{c}$ are the hopping amplitudes along the chains and along the $c$ axis, respectively, as used in the tight-binding model (see text). (b) Fermi surface obtained at 8~K with 31~eV photon energy in the $b^{*}c^{*}$ plane. The rectangle shows the first Brillouin zone as in (c). Dotted lines are guides to the eye. (c) DFT Fermi-surface contours in the $b^{*}c^{*}$ plane for various momenta covering the full Brillouin zone along $a^{*}$.
}
\end{figure}

\begin{figure}[tb]
\includegraphics[width=\columnwidth]{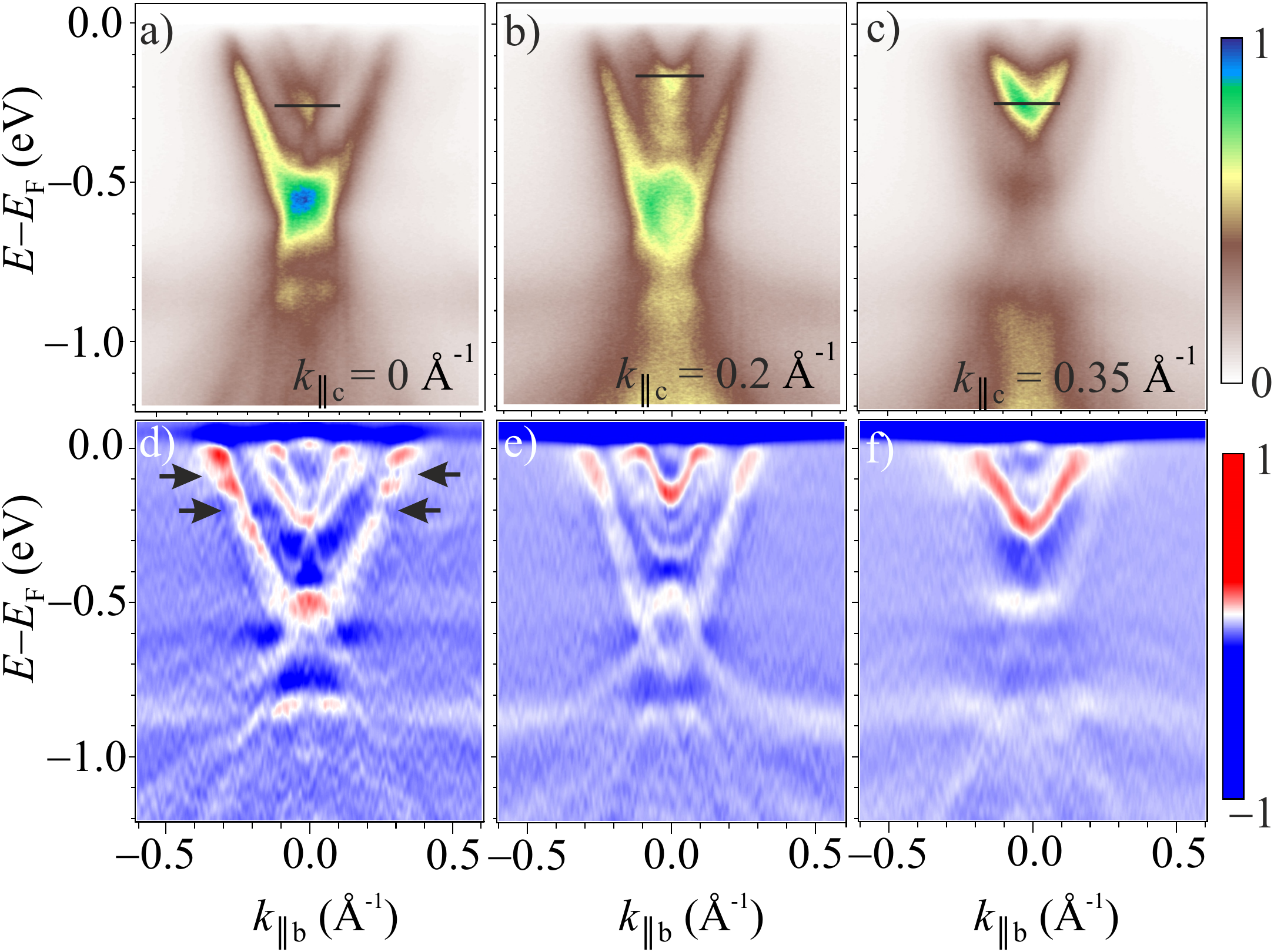}
\caption{\label{fig:Figure 2}
(Color online) (a)--(c) Band dispersion along $k_{\parallel \mathrm{b}}$ obtained at 6.5~K with 31~eV photon energy at the marked $k_{\parallel \mathrm{c}}$ values. Black bars indicate the expected position of the band bottom based on a cosine dispersion along $k_{\parallel \mathrm{c}}$ with bandwidth $4t_{c}$~=~108~meV (see text). (d)--(f) Corresponding second-derivative plots. The position of the CDW gaps are marked in (d).
}
\end{figure}

A schematic of the crystal structure of \NbSe{} is presented in Fig.~\ref{fig:Figure 1}a, and comprises three distinct triangular prism chains running parallel to the $b$ axis. The Fermi surface obtained by ARPES in the $bc$-plane is shown in Fig.~\ref{fig:Figure 1}b and may be compared with that calculated by Density Functional Theory (DFT) in Fig.~\ref{fig:Figure 1}c (calculations as in Ref. \cite{Schafer2001} using Wien2k \cite{Blaha2001}). While not all five sheets predicted by DFT are resolved at the Fermi level, Fig.~\ref{fig:Figure 2}b reveals five bands dispersing up to $E_{\mathrm{F}}$ in agreement with the DFT predictions (see Supplemental Material for further details, and out of plane dispersion including Ref.~\cite{Damascelli2004}). The dispersion is strongly anisotropic, revealing the quasi-1D nature of the electronic structure. Warping of the Fermi-surface sheets along the $k_{\parallel \mathrm{c}}$ direction (along the $c$-axis in real space) reveals the presence of significant inter-chain coupling at these low temperatures.

The dispersion of the bands along $k_{\parallel \mathrm{b}}$ at selected $k_{\parallel \mathrm{c}}$ values is given in Figs.~\ref{fig:Figure 2}a--c. Second derivative plots are presented in Figs.~\ref{fig:Figure 2}d--f in order to highlight weak features. A number of features are visible which were not resolved in previous studies \cite{Schafer2001,Schafer2003a}. At $k_{\parallel \mathrm{c}}=0$~\AA$^{-1}$ we observe three bands dispersing symmetrically around the $\Gamma$ point: the outer band with minimum at $-550$~meV and two inner bands with minima at $-260$~meV. All three bands appear to cross $E_{\mathrm{F}}$, although the spectral weight strongly decreases at low binding energies. A very small pocket directly at $E_{\mathrm{F}}$ around $\Gamma$ is also observed. At $k_{\parallel \mathrm{c}}=0.2$~\AA$^{-1}$ three inner bands can be distinguished. At $k_{\parallel \mathrm{c}}=0.35$~\AA$^{-1}$, in the second Brillouin zone, strong effects of the varying photoemission matrix element lead to different relative intensities of the bands.

\begin{figure}[tb]
\includegraphics[width=\columnwidth]{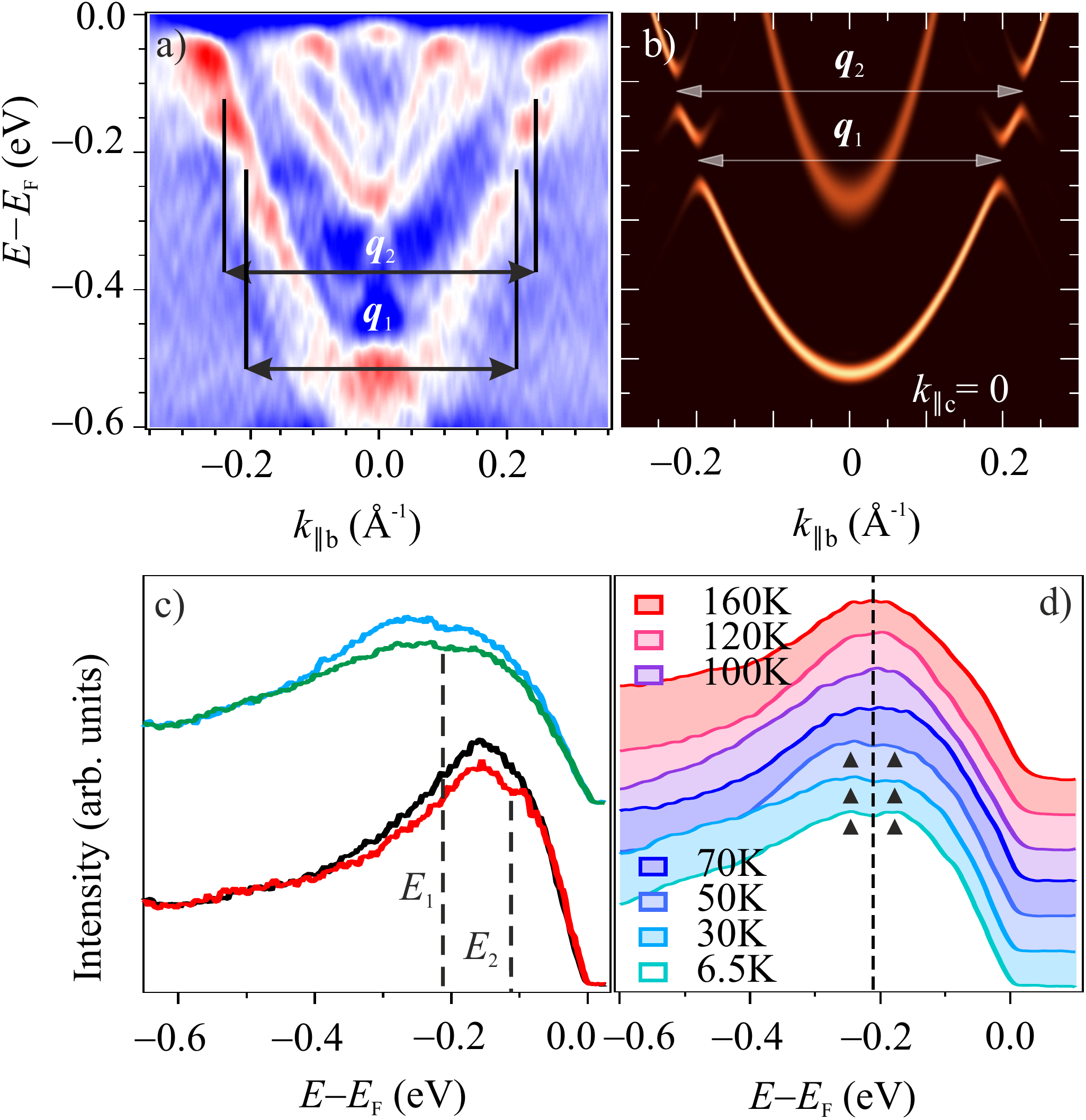}
\caption{\label{fig:Figure 3}
(Color online) (a) Zoom of the data from Fig.~\ref{fig:Figure 2}d including the $b^{*}$-axis component of the $\vec{q}$ vectors for the two incommensurate CDWs: $q_{1,b^{*}}=0.435$~\AA$^{-1}$; $\vec{q}_{2,b^{*}}=0.468$~\AA$^{-1}$ as measured by x-ray diffraction \cite{Fleming1978, Hodeau1978}. (b) Simulation of the spectral function in a CDW system with ordering vectors $q_{1,b^{*}}$ and $q_{2,b^{*}}$ (see text). (c) EDCs of the raw ARPES data at $k_{\parallel \mathrm{b}}=0.21$~\AA$^{-1}$ (blue), $k_{\parallel \mathrm{b}}=-0.20$\AA~$^{-1}$ (green), $k_{\parallel \mathrm{b}}=0.25$~\AA$^{-1}$ (red) and $k_{\parallel \mathrm{b}}=-0.24$~\AA$^{-1}$ (black). The center of the CDW gaps is marked by dashed lines. All data are taken at $k_{\parallel \mathrm{c}}=0$ and $T=6.5$~K. (d) Temperature dependence of the $\vec{q}_1$ gap at $k_{\parallel \mathrm{b}}=0.21$~\AA$^{-1}$.
}
\end{figure}

Fig.~\ref{fig:Figure 2}d further reveals a loss of intensity in the outer band at specific energies which appear symmetrically on both sides of $\Gamma$ (arrows), emphasised in Fig.~\ref{fig:Figure 3}a. To ensure these are not artefacts of the image processing, we present in Fig.~\ref{fig:Figure 3}c energy distribution curves (EDCs) of the raw data at the momenta corresponding to these features. Weak but distinct two-peak structures are observed, centered around $E_{1}=210$~meV and $E_{2}=120$~meV below $E_{\mathrm{F}}$, which we identify with gaps caused by the $\vec{q}_1$ and $\vec{q}_2$ CDW super-periodicities, respectively. The scattering vectors deduced from our data, 0.43~\AA$^{-1}$ and 0.47~\AA$^{-1}$, match within errors the $b^{*}$ components of the $\vec{q}_{1}$ and $\vec{q}_{2}$ modulation obtained by x-ray diffraction \cite{Fleming1978}. This is strong evidence that these gap features occur as a result of the CDWs \footnote{The formation of both gaps in the outer band and the observation of both modulations on all chains by STM \cite{Brun2009, *Brun2010} show that Nb orbitals on the three types of chains are coupled and feel both CDW potentials.}. In addition, as the temperature is increased, the gap features become weaker up to 50~K and then disappear, as shown for the $\vec{q}_1$ gap in Fig.~\ref{fig:Figure 3}d. The fact that the CDW gaps disappear before the bulk transition temperature should not be taken as evidence for a lower CDW transition temperature at the surface, but more likely due to phonon broadening washing out the signal as $T$ increases. The occurrence of the gap at 210~meV agrees with that observed in Refs.~\onlinecite{Schafer2001, Schafer2003a}, although the dispersion that was extracted is not reproduced in our data.

A calculation of the spectral function for a two-band tight-binding model with a two-component CDW is presented in Fig.~\ref{fig:Figure 3}b (see Supplemental Material for details including Refs.~\cite{Covaci2010,Weisse2006,Prodan1996}). The CDW potential opens gaps at momenta connected by the ordering vectors. As is evident from this simulation, the influence of the CDW on the spectral function can be rather weak and the inner band is not affected at all.

We note that the shape of the innermost band in Fig.~\ref{fig:Figure 2}e appears to bend away from the Fermi level at $k_{\parallel \mathrm{b}}=0.11$~\AA$^{-1}$, consistent with previous observations which assigned this behavior to the $\vec{q}_{2}$ CDW \cite{Schafer2003a}. We caution that such behavior may result from artifacts due to the second-derivative image processing in the presence of multiple bands and the Fermi edge. The energy and momentum distribution curves presented in the Supplemental Material confirm the absence of a backfolded dispersion. While we cannot rule out gaps at $E_{\mathrm{F}}$ that are hidden by the depletion of spectral weight, our data and calculations reveal that the CDW wave vectors $\vec{q}_{1}$ and $\vec{q}_{2}$ open gaps only below $E_{\mathrm{F}}$, which speaks against a Fermi-surface instability.

\begin{figure}[t]
\includegraphics[width=\columnwidth]{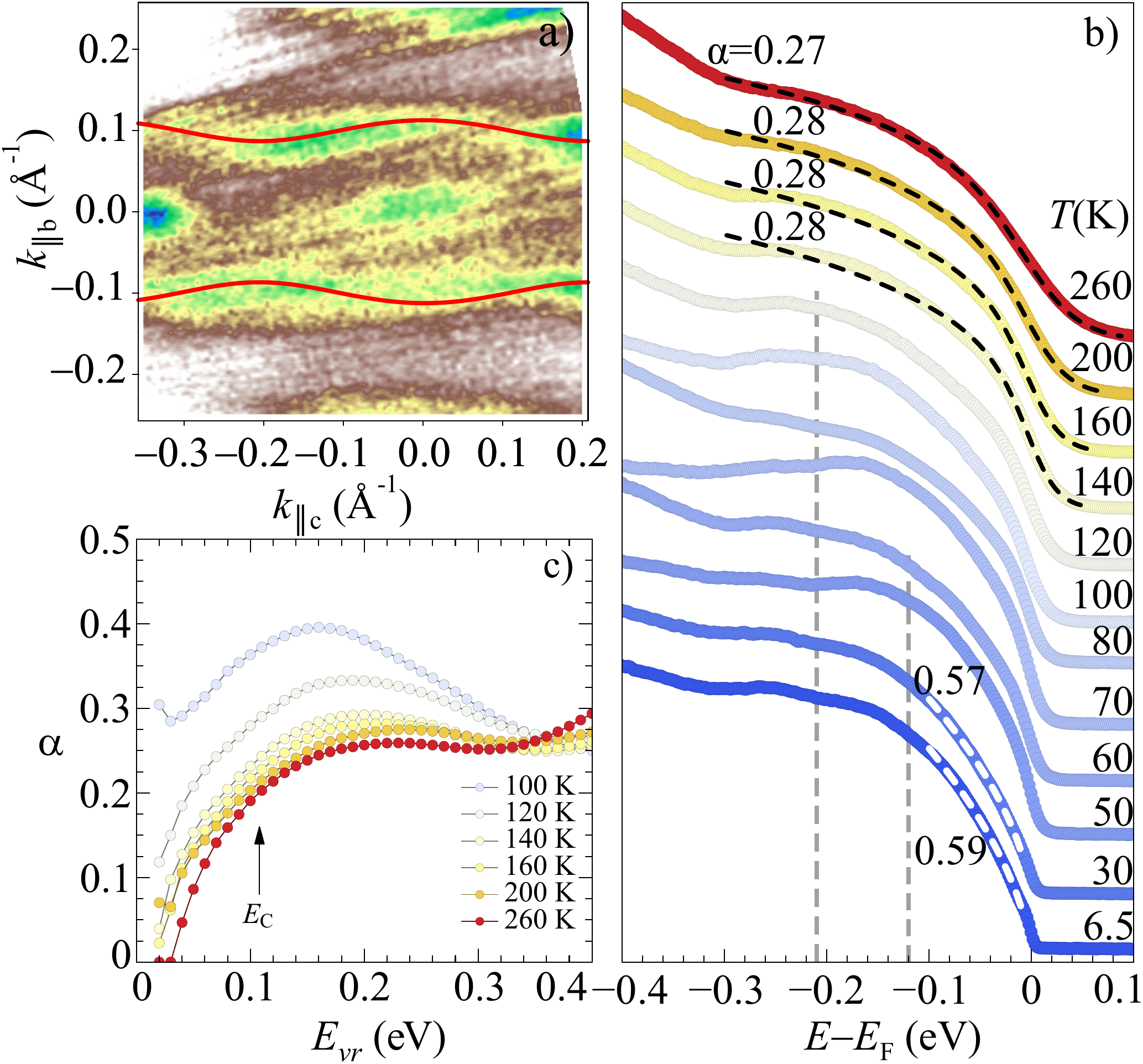}
\caption{\label{fig:Figure 4}
(Color online) (a) Fermi surface obtained at $h\nu=22$~eV, overlaid with the tight-binding model described in the text. (b) $k_{\parallel \mathrm{b}}$-integrated ARPES intensity at $k_{\parallel \mathrm{c}}=0$ for various temperatures. White dashed lines indicate simple power laws with exponents close to 0.6; black dashed lines show best fits to a TLL model spectral function with $\alpha\approx0.25$; vertical dashed lines indicate the energies of the CDW gaps seen in Fig.~\ref{fig:Figure 2}d. (c) Energy-dependent exponent showing dimensional crossover at $E_{\mathrm{C}}$ for $T>T_1$.
}
\end{figure}

We now look in more detail at the Fermi-surface data presented in Fig.~\ref{fig:Figure 4}a. It is clear that the dispersion in the $(k_{\parallel \mathrm{b}},k_{\parallel \mathrm{c}})$ plane is strongly anisotropic (i.e.\ quasi-1D), with a finite warping along $k_{\parallel \mathrm{c}}$ resulting from interchain hopping. Such a quasi-1D dispersion is minimally described by the tight-binding model 
	\begin{equation}\label{eq:band}
		E_{\vec{k}}=-2t_{b}\cos(k_{\parallel \mathrm{b}}b)-2t_{c}\cos(k_{\parallel \mathrm{c}}c)-\mu,
	\end{equation}
in which $t_{b}$ and $t_{c}\ll t_{b}$ are the effective hopping amplitudes along the chains and along the $c$ axis, respectively, and $\mu$ is the chemical potential. The transverse bandwidth $4t_{c}\equiv E_{\mathrm{C}}$ defines the energy (temperature) scale at which the system crosses over from 1D to higher dimensional behavior. Excitations with energies $E\gg E_{\mathrm{C}}$ are insensitive to the dispersion along $c$ and exhibit 1D character, with the typical power laws expected for a TLL, while excitations with $E<E_{\mathrm{C}}$ behave as in a FL. Our Fermi-surface and dispersion data (Figs.~\ref{fig:Figure 4}a and Fig.~SM2 of Supplemental Material) in combination with Eq.~(\ref{eq:band}) allow us to extract a value $t_{c}=27$~meV (for a derivation see Supplemental Material). This implies a typical energy scale $E_{\mathrm{C}}=108$~meV above which 1D signatures should be observed. A simple cross check of this value can be obtained by comparing the relative band bottoms throughout the Brillouin zone along the $k_{\parallel \mathrm{c}}$ direction with that expected from a bandwidth of $4t_{c}$. These are presented in Fig.~\ref{fig:Figure 2}a-c by black horizontal bands and show good agreement.

In Fig.~\ref{fig:Figure 4}b we show the $k_{\parallel \mathrm{b}}$-integrated ARPES intensity at $k_{\parallel \mathrm{c}}=0$ within 400~meV of $E_{\mathrm{F}}$ and for different temperatures. This quantity approximates the density of states (DOS) multiplied by the Fermi function, apart from inessential corrections associated with the finite experimental resolution and weak $k_{\parallel \mathrm{c}}$ dispersion. We have checked that integrating over smaller $k$-ranges or at other $k_{\parallel \mathrm{c}}$ values does not change the form of the DOS, see Supplemental Material. At the lowest temperatures, the DOS suppression near $E_{\mathrm{F}}$ is markedly different from the expected Fermi edge and resembles a power law. We rule out CDW gaps as a possible explanation for this anomalous suppression of spectral weight: the CDW gaps seen at finite energy in our data have a typical peak-to-peak size of 70~meV (Fig.~\ref{fig:Figure 3}c), while the DOS suppression occurs over a much wider energy range. A complete gapping of the Fermi surface would also be inconsistent with the fact that \NbSe{} remains metallic even at these low temperatures \cite{Monceau1977}.

The power-law depletion evokes a TLL where the DOS is suppressed like $|E-E_{\mathrm{F}}|^{\alpha}$ at low energy due to the disappearance of single-particle excitations \cite{Voit1994}. The non-universal exponent $\alpha=(K_{\rho}+K_{\rho}^{-1}-2)/4$ (see Supplemental Material) depends on the parameter $K_{\rho}$ which measures the strength of interactions and varies between 0 $< K_{\rho} \leqslant$ 1. $K_{\rho}=1$ ($\alpha=0$) corresponds to a non-interacting electron system with flat DOS.

TLL signatures should be searched for at energies and/or temperatures larger than $E_{\mathrm{C}}$. Our highest measured temperature (260~K) is well below the crossover scale ($E_{\mathrm{C}} \approx 110$ meV $\approx 1250$~K), such that our whole data set can be regarded as being in a low-temperature regime with respect to the dimensional crossover. In the CDW state, the DOS at $E>E_{\mathrm{C}}$ is perturbed by the CDW gaps such that TLL signatures may be masked if present. We therefore look for the TLL power law at temperatures higher than $T_1=145$~K. We fit the data with an expression giving the finite-$T$ DOS of a TLL \cite{Schonhammer1993} convolved by our experimental resolution (for further details see Supplemental Material). The fits are performed in a variable energy range $[-E_{\mathrm{vr}},\min(E_{\mathrm{vr}},4k_{\mathrm{B}}T)]$ around $E_{\mathrm{F}}$ and we extract the exponent $\alpha$ as a function of this range (Fig.~\ref{fig:Figure 4}c). At $E_{\mathrm{vr}}<E_{\mathrm{C}}$, the data approach a pure Fermi edge with $\alpha=0$ (3D regime), while at $E_{\mathrm{vr}}>E_{\mathrm{C}}$, the fit yields a stable exponent $\alpha\approx0.25$ ($K_{\rho}=0.38$) over a broad energy range (1D regime). The fit includes all data at $E<E_{\mathrm{vr}}$ and therefore yields a continuous drop of $\alpha$ towards zero when reducing $E_{\mathrm{vr}}$ below $E_{\mathrm{C}}$; this trend is observed at temperatures above 120~K. An analysis of the fit quality (see Supplemental Material) shows that the best fits are obtained for an upper $E_{\mathrm{vr}}$-bound between $0.2$ and $0.3$~eV. Beyond this the DOS upturn from $-0.3$~eV due to the band bottom at $-0.6$~eV means that the power-law analysis in this range is no longer appropriate. Figure~\ref{fig:Figure 4}b shows the best fits with $E_{\mathrm{vr}}=0.3$~eV and the corresponding TLL exponents for $T>100$~K. For $T<T_{1}$, the fit deteriorates and the extracted exponent becomes strongly energy dependent due to the CDW gaps, while the exponent at low $E_{\mathrm{vr}}$ increases steadily with decreasing $T$ due to the anomalous spectral-weight suppression. 
The high-$T$ value $K_{\rho}=0.38$ suggests that the interaction has a finite range. Indeed $K_{\rho}>1/2$ in the Hubbard model while $K_{\rho}>1/8$ in the extended Hubbard model \cite{Voit1994}. The value 0.38 thus points to a moderate interaction and locates NbSe$_3$ far from an interaction-driven metal-insulator transition. In contrast, the apparent DOS exponent close to 0.6 at low $T$ (Fig.~\ref{fig:Figure 4}b) would indicate much stronger correlations with $K_{\rho}=0.24$. Since our DFT Fermi surface and bands agree with the observed band structure it is unlikely that such a strong renormalization occurs. We conclude that the TLL spectral function is not an appropriate description of the system at these low temperatures where a FL phase is expected. We note that DOS exponents close to 0.6 have been reported for several systems, which may indicate longer-range interactions as in carbon nanotubes \cite{Ishii2003} and atomic chains \cite{Meyer2014, Blumenstein2011, Ohtsubo2015} or multiband effects like in lithium purple bronze \cite{Wang2006, Dudy2013}.

Our analysis supports the idea that \NbSe{} is never observed in a truly TLL regime up to room temperature. Instead a gradual crossover from 1D to 3D occurs, as evidenced by the energy-dependence of the $\alpha$ exponent. Such a dimensional crossover is expected in all quasi-1D materials at the energy of the renormalized interchain coupling. This invites to reconsider previous reports of TLL power-law DOS suppression and check whether the exponents were indeed measured in the 1D regime where the analysis is valid.

In summary, we have performed detailed ARPES measurements over a wide temperature range which reveal evidence of a dimensional crossover in \NbSe{}. Such a dimensional crossover is consistent with the quasi-1D warping of the Fermi surface. A careful analysis of the density of states and comparison with expectations for 1D behavior reveal a changing dimensionality of excitations above a characteristic energy $E_{\mathrm{C}}$. At low temperatures we observe CDW gaps in the electronic structure at the momenta indicated by x-ray diffraction. We expect the analysis presented here to be applicable to other quasi-1D systems due to the generality of finite inter-chain coupling in real materials, and hope this will stimulate further experimental and theoretical research on the dimensional crossover.

\begin{acknowledgments}

We thank Diamond Light Source for access to beamline I05 (SI10322) and acknowledge useful discussions with Y. Ohtsubo and A. P. Petrovi\'{c}. The work done in Geneva (C.B.) was supported by the Swiss National Science Foundation under Division II. C.M. gratefully acknowledges the support of the SNSF under grant $PZ00P2\_ 154867$.

\end{acknowledgments}


%

\onecolumngrid
\newpage
\begin{center}
{\large\textbf{\boldmath
Supplemental Material\\ [0.5em] {\small to} \\ [0.5em]
Dimensional Crossover in a Charge Density Wave Material Probed\\[0.2em] by Angle-Resolved Photoemission Spectroscopy
}}\\[1.5em]

C. W. Nicholson,$^1$ C. Berthod,$^2$ M. Puppin,$^1$ H. Berger,$^3$ M. Wolf,$^1$ M. Hoesch,$^4$, and C. Monney$^5$\\[0.5em]

\textit{\small
$^1$Department of Physical Chemistry, Fritz-Haber-Institut of the Max Planck Society, Faradayweg 4-6, Berlin 14915, Germany\\
$^2$Department of Quantum Matter Physics, University of Geneva,
24 quai Ernest-Ansermet, 1211 Geneva, Switzerland\\
$^3$Institut de la Mati\`ere Complexe,
\'Ecole Polytechnique F\'ed\'erale de Lausanne, 1015 Lausanne, Switzerland\\
$^4$Diamond Light Source, Harwell Campus, Didcot OX11 0DE, Oxfordshire, United Kingdom\\
$^5$Institute of Physics, University of Zurich, Winterthurerstrasse 190, 8057 Zurich, Switzerland
}

\vspace{2em}
\end{center}

\twocolumngrid
\setcounter{figure}{0}
\setcounter{equation}{0}
\renewcommand{\theequation}{SM\arabic{equation}}
\renewcommand{\thefigure}{SM\arabic{figure}}
\renewcommand{\thetable}{SM\arabic{table}}

\section*{Determination of tight-binding parameters}

\begin{figure}[b]
\includegraphics[width=\columnwidth]{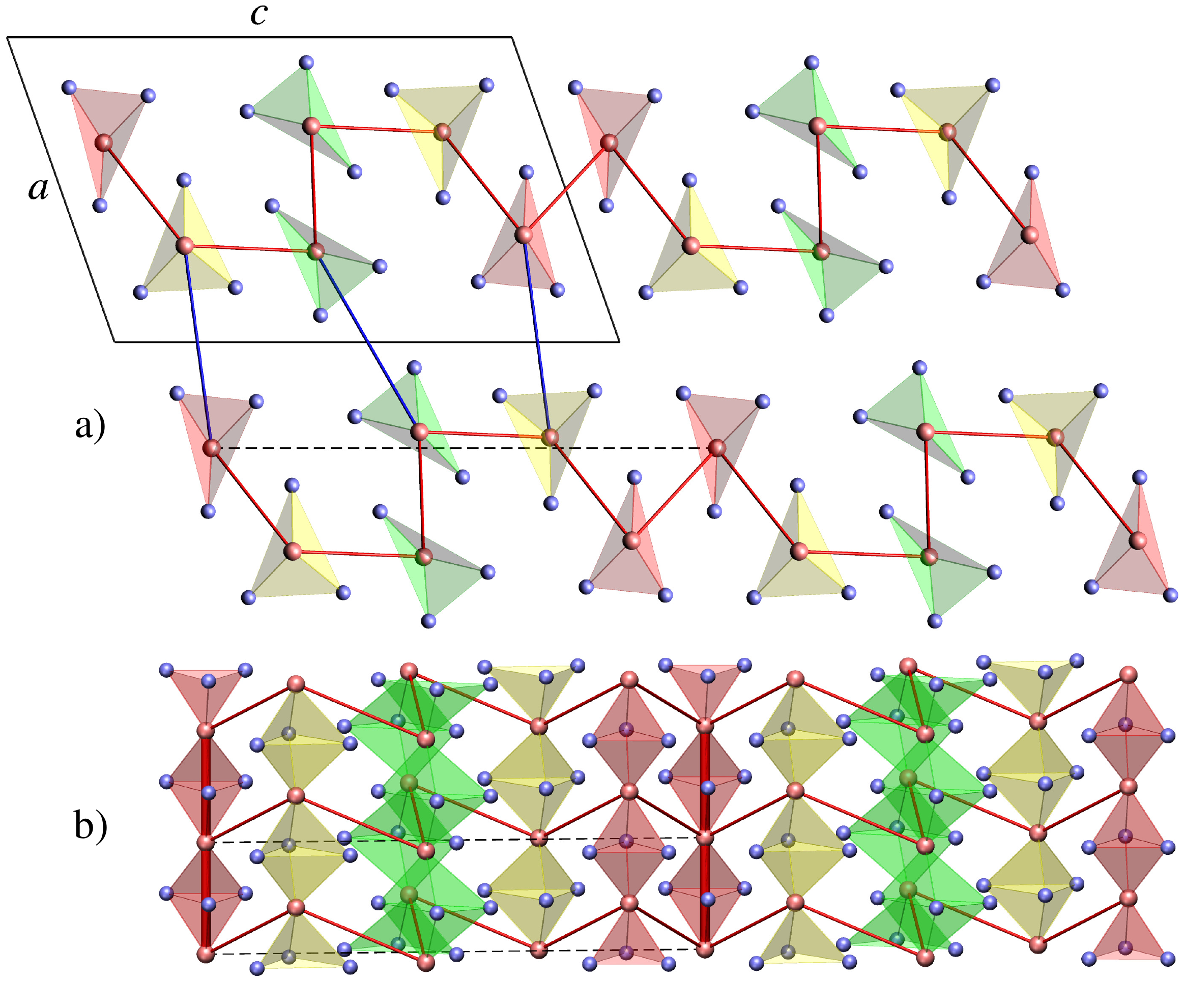}
\caption{\label{fig:structure}
(a) Top view ($ac$ plane) and (b) side view ($bc$ plane) of the \NbSe{} crystal with Nb atoms in red and Se atoms in blue. The unit cell contains six chains running along the $b$ axis, forming three pairs related by inversion symmetry and highlighted in red (denoted I in Fig.~1a of the main text), yellow (II), and green (III). 
Solid lines: unit cell in the $ac$ plane; dashed lines: unit cell in the $bc$ plane; thick red cylinders: strong bonds with $t_b$ overlap integral along the chains; thin red cylinders: hopping path along the $c$ axis leading to the effective $t_{c}$ amplitude; thin blue cylinders: weak bonds along the $a$ axis.
}
\end{figure}

We describe the low-energy dispersion and the Fermi surface using the tight-binding model
	\begin{equation}\label{eq:tight-binding}
		E_{\vec{k}}=-2t_{b}\cos(k_{\parallel \mathrm{b}}b)-2t_{c}\cos(k_{\parallel \mathrm{c}}c)-\mu,
	\end{equation}
where $t_{b}$ and $t_{c}$ are the energies associated with hopping along and across the 1D chains, respectively, $b=3.48$~\AA\ and $c=15.56$~\AA\ are the lattice constants along the corresponding directions, and $\mu$ is the chemical potential. This is a minimal effective model which does not take full account of the \NbSe{} unit-cell structure. The latter involves six formula units per cell as depicted in Fig.~\ref{fig:structure}. The dispersion along the $c$ axis implies hopping through six Nb--Nb bonds running across the unit cell, leading to an effective hopping amplitude $t_c$ from one unit cell to the next. The length of these bonds varies between 4.2 and 4.4~\AA, slightly longer than the strong Nb--Nb bonds along $b$. The dispersion along the $a$ axis is more strongly suppressed due to larger distances between the Nb atoms (6.5--6.6~\AA).

\begin{figure}[b]
\includegraphics[width=\columnwidth]{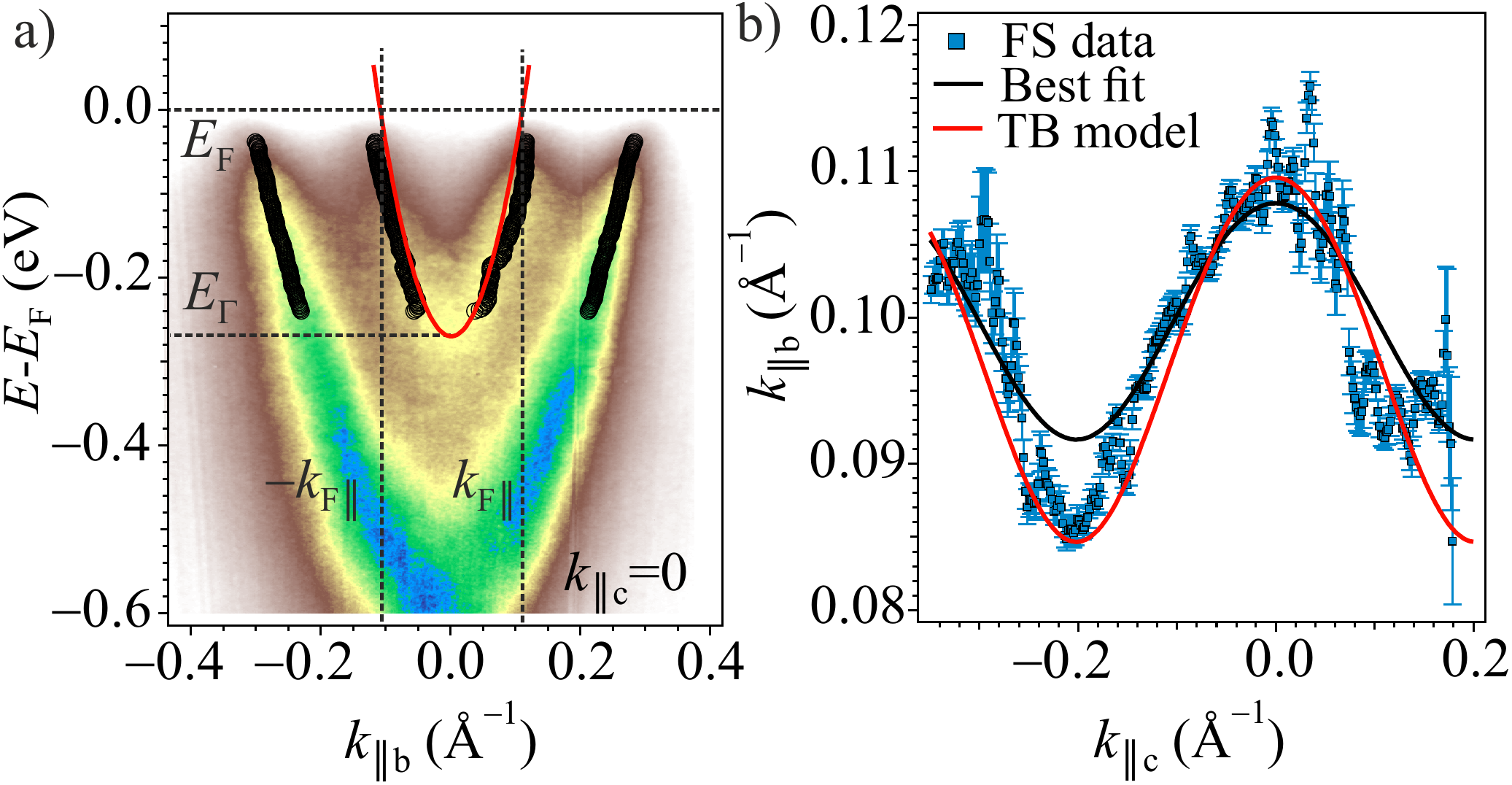}
\caption{\label{fig:dispersion}
(a) ARPES intensity vs $E-E_{\mathrm{F}}$ and $k_{\parallel\mathrm{b}}$ at $k_{\parallel\mathrm{c}}=0$, with the position of the bands extracted from the MDCs overlaid in black. The minimum and Fermi points are shown for the inner band. The solid red line is the tight-binding model discussed in the text. (b) Position of the inner-band Fermi surface (upper branch) extracted from MDC fits overlaid with the tight-binding model (red) and a Fermi-surface fit (black).
}
\end{figure}

The data that we have used in order to determine the tight-binding parameters of the model are presented in Fig.~\ref{fig:dispersion}. We focus on the inner band, where the Fermi-surface warping is most clearly observed. Fig.~\ref{fig:dispersion}a shows the dispersion at $k_{\parallel\mathrm{c}}=0$ and Fig.~\ref{fig:dispersion}b shows the Fermi surface of the inner band around $k_{\parallel\mathrm{b}}=0.1~\text{\AA}^{-1}$. The Fermi surface positions are extracted from the data of Fig.~4a in the main text by fitting vertical cuts through the 2D data set. We then perform a cosine fit of the extracted Fermi surface, in order to determine the tight-binding parameters. However, as the Fermi surface is invariant under an arbitrary scaling of the whole dispersion, such a fit does not allow one to determine the absolute values of the parameters, but only relative values. We thus obtain $t_c/t_b=0.00956$ and $\mu/t_b=-1.88$, corresponding to the black curve in Fig.~\ref{fig:dispersion}b. This curve shows that this best fit model is too crude to capture all details of the $c$-axis dispersion; in particular, it severely underestimates the warping of the Fermi surface. As an effective model, we expect (\ref{eq:tight-binding}) to reproduce at least the bandwidth along $k_{\mathrm{\parallel b}}$ and the warping of the Fermi surface. Since the best fit shown in Fig.~\ref{fig:dispersion}b does not provide absolute values and furthermore underestimates the warping, we proceed differently and determine the three tight-binding parameters from the conditions that (i) the energy at the band bottom for $k_{\parallel\mathrm{c}}=0$ is $E_{\Gamma}=-0.262$~eV as shown in Fig.~\ref{fig:dispersion}a; (ii) the Fermi wave vector at $k_{\parallel \mathrm{c}}=0$ is $k_{\mathrm{F}\parallel\mathrm{b}}=0.108$~\AA$^{-1}$ as observed in Fig.~\ref{fig:dispersion}b; and (iii) the Fermi-surface warping is given by the peak-to-peak value observed in Fig.~\ref{fig:dispersion}b, namely $\Delta k_{\parallel\mathrm{b}}=0.0247$~\AA$^{-1}$.

At $k_{\parallel\mathrm{c}}=k_{\parallel\mathrm{b}}=0$, the expression (\ref{eq:tight-binding}) simplifies to
	\begin{equation}\label{eq:Gamma}
		E_{\Gamma}=-2t_{b}-2t_{c}-\mu.
	\end{equation}
The Fermi point at $k_{\parallel \mathrm{c}}=0$ satisfies the equation
	\begin{equation}\label{eq:kFy}
		0=-2t_{b}\cos(k_{\mathrm{F}\parallel\mathrm{b}}b)-2t_{c}-\mu.
	\end{equation}
Inserting the values of $E_{\Gamma}$, $k_{\mathrm{F}\parallel\mathrm{b}}$, and $b$ in Eqs.~(\ref{eq:Gamma}) and (\ref{eq:kFy}), we obtain $t_{b}=1.88$~eV and $\mu=-3.49~\mathrm{eV}-2t_{c}$. Next we make use of the Fermi-surface warping. The extremal values of $k_{\parallel\mathrm{b}}$ occur for $\cos(k_{\parallel \mathrm{c}}c)=\pm1$. Rearranging Eq.~(\ref{eq:tight-binding}) for $E_{\vec{k}}=0$, we therefore arrive at:
	\begin{equation}\label{eq:warping}
		\Delta k_{\parallel\mathrm{b}}=\frac{1}{b}\left[
		\cos^{-1}\left(\frac{\mu-2t_{c}}{2t_{b}}\right)
		-\cos^{-1}\left(\frac{\mu+2t_{c}}{2t_{b}}\right)\right].
	\end{equation}
Substituting the known values and the relation $\mu=-3.49~\mathrm{eV}-2t_c$ leads to the solution $t_c=0.027$~eV. The resulting tight-binding dispersion is displayed in Figs.~\ref{fig:dispersion}a and \ref{fig:dispersion}b as a solid red line. This set of tight-binding parameters gives a better account of the Fermi-surface warping than the Fermi-surface fit.

\section*{Comparison with DFT}

\begin{figure}[t]
\includegraphics[width=\columnwidth]{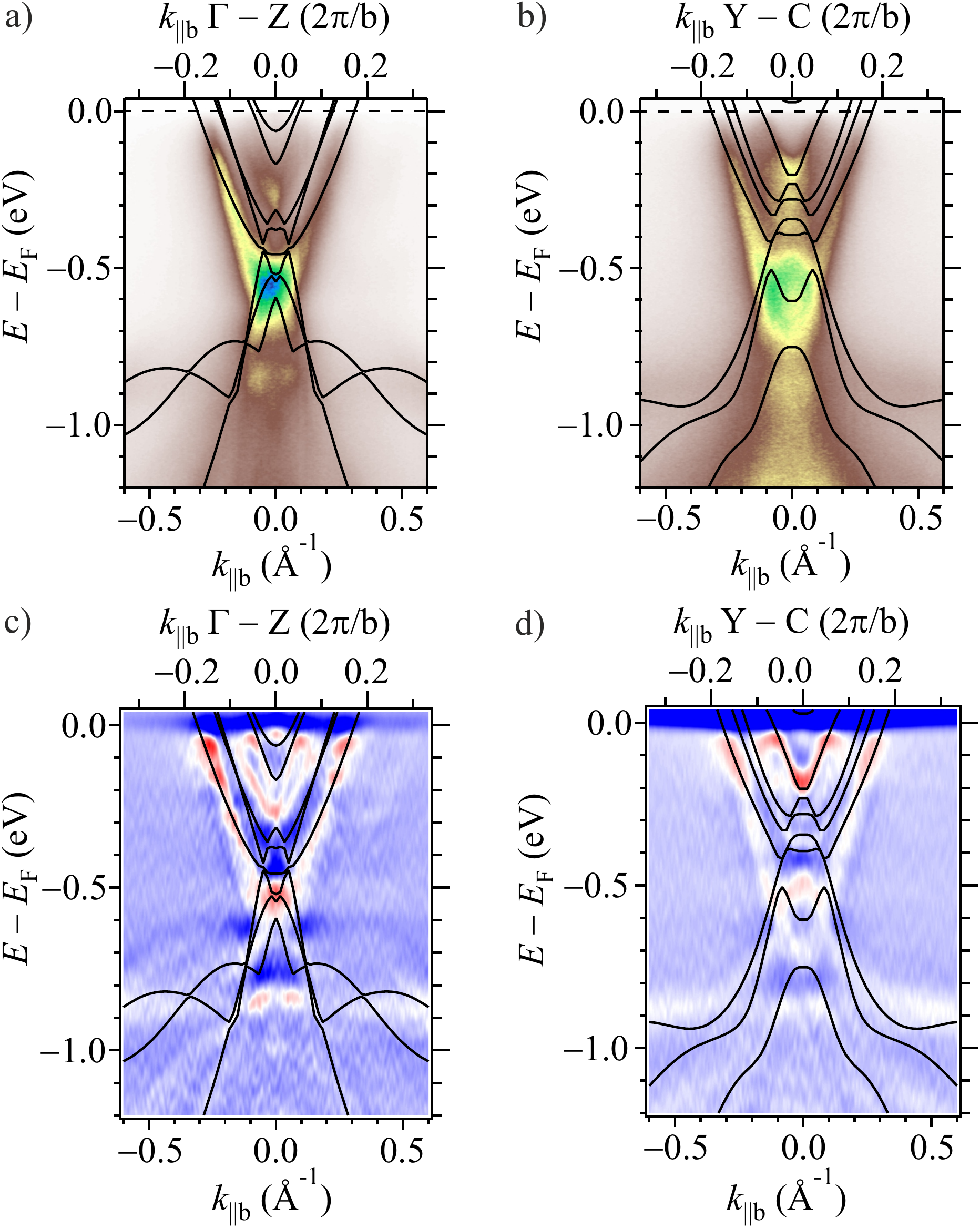}
\caption{\label{fig:DFT}ARPES data and the second derivative along $\Gamma$--Z (a) and (c) and Y--C (b) and (d), as shown in Fig.~2 of the main text, overlaid with DFT calculations along the respective high-symmetry lines.
}
\end{figure}

In Fig.~\ref{fig:DFT}, we present a comparison of the ARPES band structure (already shown in Fig.~2 of the main text) with the band structure obtained from DFT calculations. Cuts are shown along the $\Gamma$--Z and Y--C directions. Both the positions and band widths of the experimental bands are in overall good agreement with the DFT, which points to minimal renormalization effects as a result of e.g.\ electron-electron interactions. Our calculations indicate that all bands close to the Fermi level have predominantly Nb $d$-orbital character. The Fermi wave vectors of experiment and DFT are compared in Table~\ref{tab:kF} and confirm the good agreement. The five bands crossing the Fermi level in the DFT calculation are numbered 1--5 from outer to inner.

\begin{table}[h]
\caption{Comparison of Fermi wave vectors (\AA$^{-1}$) obtained from ARPES and DFT along two high-symmetry directions.}
\label{tab:kF}
\begin{tabular*}{\columnwidth}{@{\extracolsep{\fill}}lcccc}
\hline \hline
	& \multicolumn{2}{c}{$\Gamma$--Z} & \multicolumn{2}{c}{Y--C} \\
     & Expt. & DFT & Expt. & DFT \\ \hline
$k_{\mathrm{F \parallel\mathrm{b}}}^{1}$ & 0.27(3)  & 0.3        & 0.27(5)    & 0.31       \\
$k_{\mathrm{F \parallel\mathrm{b}}}^{4}$ & 0.10(6)  & 0.09        & 0.09(5)    & 0.12       \\
$k_{\mathrm{F \parallel\mathrm{b}}}^{5}$ & 0.00(4)  & 0.07       & 0.00(6)    & Above $E_{\mathrm{F}}$   \\ \hline \hline
\end{tabular*}
\end{table}

\section*{Out of plane dispersion}

\begin{figure}[t]
\includegraphics[width=\columnwidth]{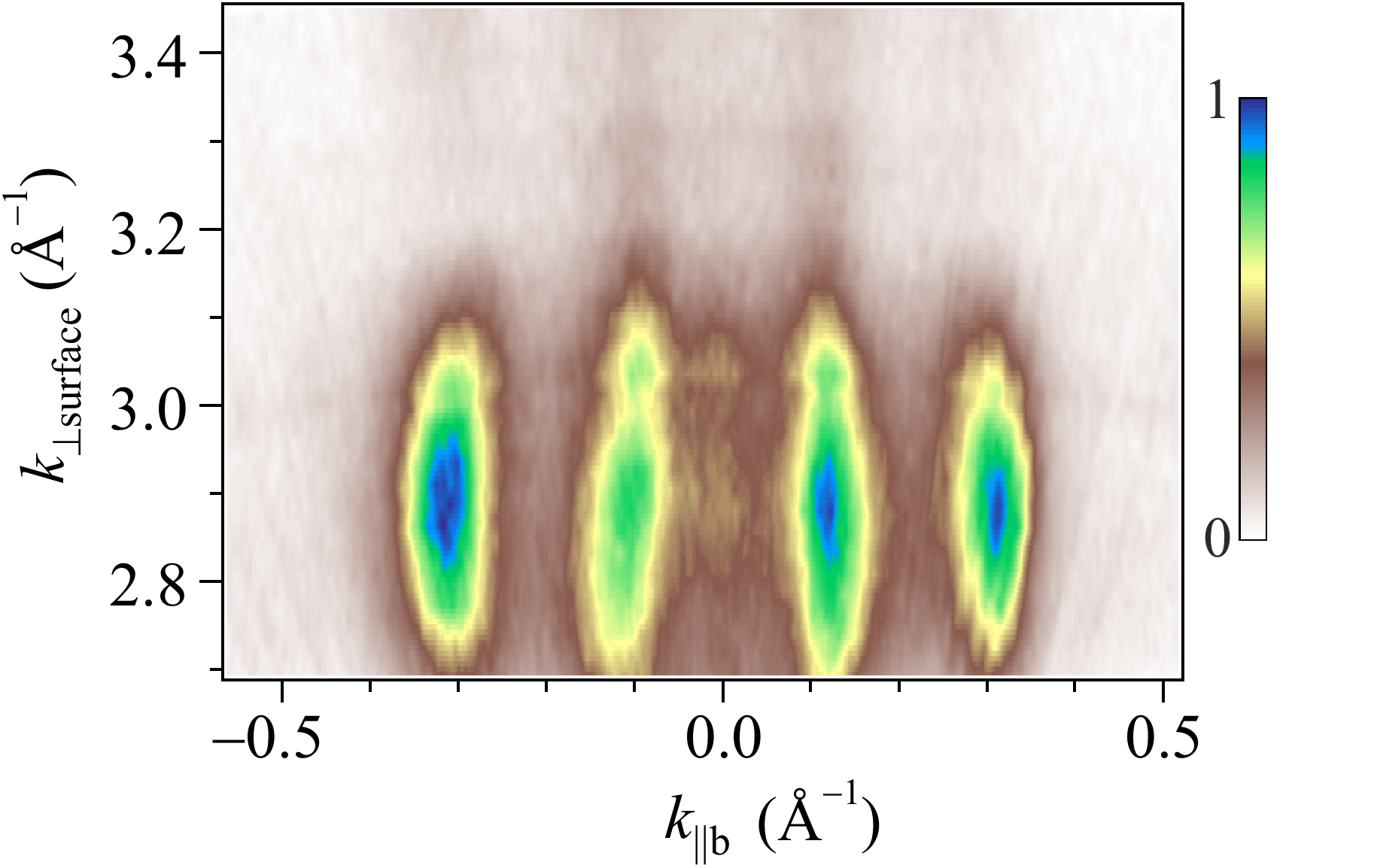}
\caption{\label{fig:kZ}
Dispersion of the Fermi surface in the plane perpendicular to the sample surface defined by the $bc$-plane.}
\end{figure}

The dispersion of the Fermi surface normal to the sample surface is presented in Fig.~\ref{fig:kZ}. This is obtained by a scan of the photon energy, in this case between 20 and 40~eV, and using the relation
	\begin{equation}\label{eq:kz}
		k_{\perp \mathrm{surface}}=\frac{1}{\hbar}\sqrt{2m(E_{\mathrm{kin}} \mathrm{cos}^{2}\theta+V_{0})},
	\end{equation}
	where terms are defined as in Ref.~\onlinecite{Damascelli2004s}. An inner potential $V_{0}=12$~eV was assumed.
The warping of the Fermi surface reveals the quasi-1D nature of the states also in this plane. The warping in this direction is certainly not more than the warping in the $bc$-plane presented in the main text, confirming the quasi-1D nature of \NbSe{}.

\section*{\boldmath Analysis of bands close to $E_{\mathrm{F}}$}

\begin{figure}[b]
\includegraphics[width=0.95\columnwidth]{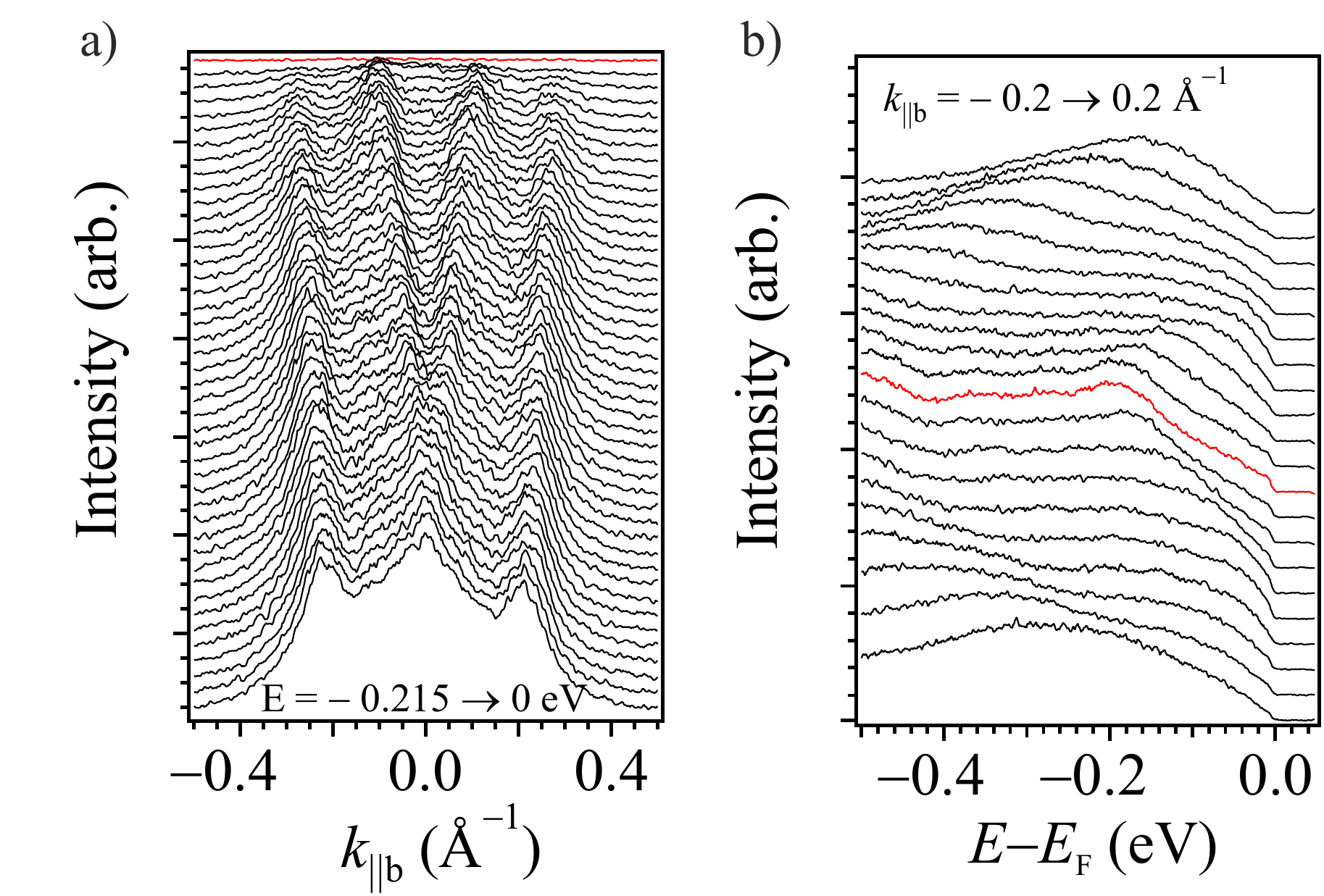}
\caption{\label{fig:DCs}
(a) MDCs and (b) EDCs of the data presented in Fig.~2b of the main text. Red curves highlight the Fermi level and $k_{\parallel \mathrm{b}}=0$, respectively.}
\end{figure}

Figure~\ref{fig:DCs} displays MDCs and EDCs close to $E_{\mathrm{F}}$ for the data presented in Fig.~2b of the main text, for which an apparent strong backfolding of the inner band is observed in the second-derivative images. As pointed out in the main text, such behavior may result from artifacts due to the second-derivative image processing in the presence of multiple bands and the Fermi edge. The energy- and momentum-distribution curves presented here confirm the absence of a backfolded dispersion.

\section*{Spectral function in a two-band quasi-1D CDW system}

We consider a two-dimensional, two-band model described by the Hamiltonian
	\begin{multline}
		H=\sum_{n\vec{k}}E^{\phantom{*}}_{n\vec{k}}c^{\dagger}_{n\vec{k}}
		c^{\phantom{\dagger}}_{n\vec{k}}
		+V_1\sum_{n\vec{k}}\left(c^{\dagger}_{n\vec{k}}
		c^{\phantom{\dagger}}_{n\vec{k}+\vec{q}_1}+\mathrm{h.c.}\right)\\
		+V_2\sum_{n\vec{k}}\left(c^{\dagger}_{n\vec{k}}
		c^{\phantom{\dagger}}_{n\vec{k}+\vec{q}_2}+\mathrm{h.c.}\right)
	\end{multline}
on a rectangular mesh with lattice parameters $b$ and $c$. The band dispersions are $E_{n\vec{k}}=-2t_{nb}\cos(k_{\parallel \mathrm{b}}b)-2t_{nc}\cos(k_{\parallel \mathrm{c}}c)-\mu_n$, $n=1,2$. With $t_{nb}\gg t_{nc}$, this represents a quasi-one dimensional lattice with main dispersion along the $y$ direction. We choose parameters that mimic the two outer bands of \NbSe: $(t_{1b},t_{1c},\mu_1)=(0.681,0.027,-0.898)$~eV and $(t_{2b},t_{2c},\mu_2)=(1.88,0.027,-3.545)$~eV. The interchain hopping $t_{nc}=27$~meV was set to match the Fermi-surface warping for band 2 (Fig.~\ref{fig:dispersion}b); we use the same value for band 1 for simplicity. The remaining parameters $t_{nb}$ and $\mu_n$ are adjusted to reproduce the Fermi points and band minima observed in Fig.~\ref{fig:dispersion}a. The last two terms in the Hamiltonian describe electrons moving in the periodic potential $V(\vec{r})=2V_1\cos(\vec{q}_1\cdot\vec{r})+2V_2\cos(\vec{q}_2\cdot\vec{r})$. This simple cosine behavior is a minimal model for a CDW with modulation vectors $\vec{q}_1$ and $\vec{q}_2$. We focus on the CDW along the chains and fix the vectors to $\vec{q}_1=[0,7\pi/(16b)]$ and $\vec{q}_2=[0,\pi/(2b)]$. We choose these values because (i) with $b=3.48$~\AA\ the wavevectors 0.395~\AA$^{-1}$ and 0.451~\AA$^{-1}$ are similar to the values 0.435~\AA$^{-1}$ and 0.468~\AA$^{-1}$ observed in Fig.~3a of the main text; (ii) these wavevectors connect $\vec{k}$ points of band 1 at energies $-209$~meV and $-119$~meV, close to the values $-210$~meV and $-120$~meV where spectral weight is suppressed in the ARPES data; and (iii) the ratio of the two wavelengths is 8/7, which leads to a commensurability with a not-too-long period of $32b$. Exact commensurability is an advantage for the calculations. The splittings of the EDC peaks due to the $\vec{q}_1$ and $\vec{q}_2$ modulations can be estimated to be 70 and 65 meV, respectively (see Fig.~3c of the main text). We therefore set the amplitudes to $V_1=70$~meV and $V_2=65$~meV.

The spectral function, to be compared with the ARPES intensity, is calculated as
	\begin{multline}
		A(\vec{k},E)=\frac{1}{S}\int d^2R
		\left(\textstyle-\frac{1}{\pi}\right)\mathrm{Im}
		\int d^2\rho\,e^{-i\vec{k}\cdot\vec{\rho}}\\
		\times G(\vec{R}+\vec{\rho}/2,\vec{R}-\vec{\rho}/2,E).
	\end{multline}
$G(\vec{r},\vec{r}',E)$ is the retarded Green's function in real space, which breaks translational invariance due to the CDW. We Fourier transform the Green's function with respect to the relative coordinate $\vec{\rho}=\vec{r}-\vec{r}'$, and perform a spatial average over the surface $S$ with respect to the center-of-mass coordinate $\vec{R}=(\vec{r}+\vec{r}')/2$. In the absence of CDW, $G$ is independent of $\vec{R}$ and the formula reduces to the usual definition for systems with translation invariance: $A(\vec{k},E)=(-1/\pi)\,\mathrm{Im}\,G(\vec{k},E)$. The real-space Green's function is calculated as $G(\vec{r},\vec{r}',E)=\langle\vec{r}|(E+i0-H)^{-1}|\vec{r}'\rangle$, by expanding $(E+i0-H)^{-1}$ on Chebyshev polynomials \cite{Covaci2010s}. The expansion is truncated to order 1000 and terminated with the Jackson kernel \cite{Weisse2006s}. The real-space system size used in the calculation contains 501'001 unit cells. The resulting spectral function shows a suppression of spectral weight at the energies that satisfy the scattering condition $E_{n\vec{k}}=E_{n\vec{k}+\vec{q}_{1,2}}$ (Fig.~3b of the main text). For the second band this condition is only met at positive energy and no signature of the CDW is therefore seen in the occupied states.

\begin{figure}[b]
\includegraphics[width=0.65\columnwidth]{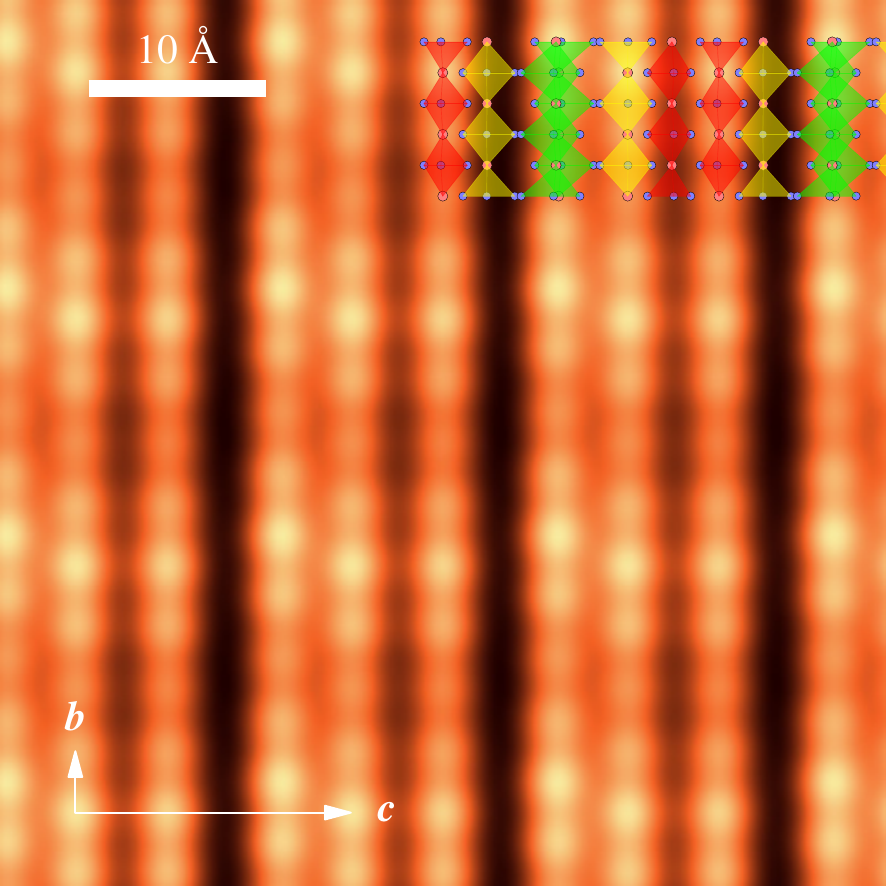}
\caption{\label{fig:STM}
Simulated STM image of the $bc$ plane of \NbSe. The projection of the crystal structure is overlaid for comparison. The CDW modulation calculated with the tight-binding model is visible along $\vec{b}$. The CDW period along $\vec{b}$ is $32b$, exceeding the size of the image.
}
\end{figure}

For a comparison with STM measurements \cite{Prodan1996s, Brun2010s} we also compute the local density of states $N(\vec{r},E)=(-1/\pi)\mathrm{Im}\,G(\vec{r},\vec{r},E)$. To mimic the STM topography, we calculate the tunneling current at $-0.3$~eV, $I(\vec{r})=\int_{-0.3~\mathrm{eV}}^0dE\,N(\vec{r},E)$, for each site of the tight-binding lattice. We then attach to each Nb atom close to the cleaving plane a Gaussian function of full width at half maximum $b$, and weight these functions according to $I(\vec{r})$. The result presented in Fig.~\ref{fig:STM} bears some resemblance with the STM data, in spite of the model simplicity. The most intense signal comes from the two type-III chains (green in Fig.~\ref{fig:structure}), one being closest to the STM tip and the second one almost exactly beneath the first one. One of the type-I chains (red) is at the surface while the second one is beneath the surface, only visible through a narrow channel, leading to a low-intensity line. Finally, the darkest regions correspond to one of the type-II (yellow) chains lying below the top Nb layer and visible through a wider channel between type-I and type-II surface chains (see Fig.~\ref{fig:structure}).

\section*{Density of states}

\begin{figure}[h]
\includegraphics[width=\columnwidth]{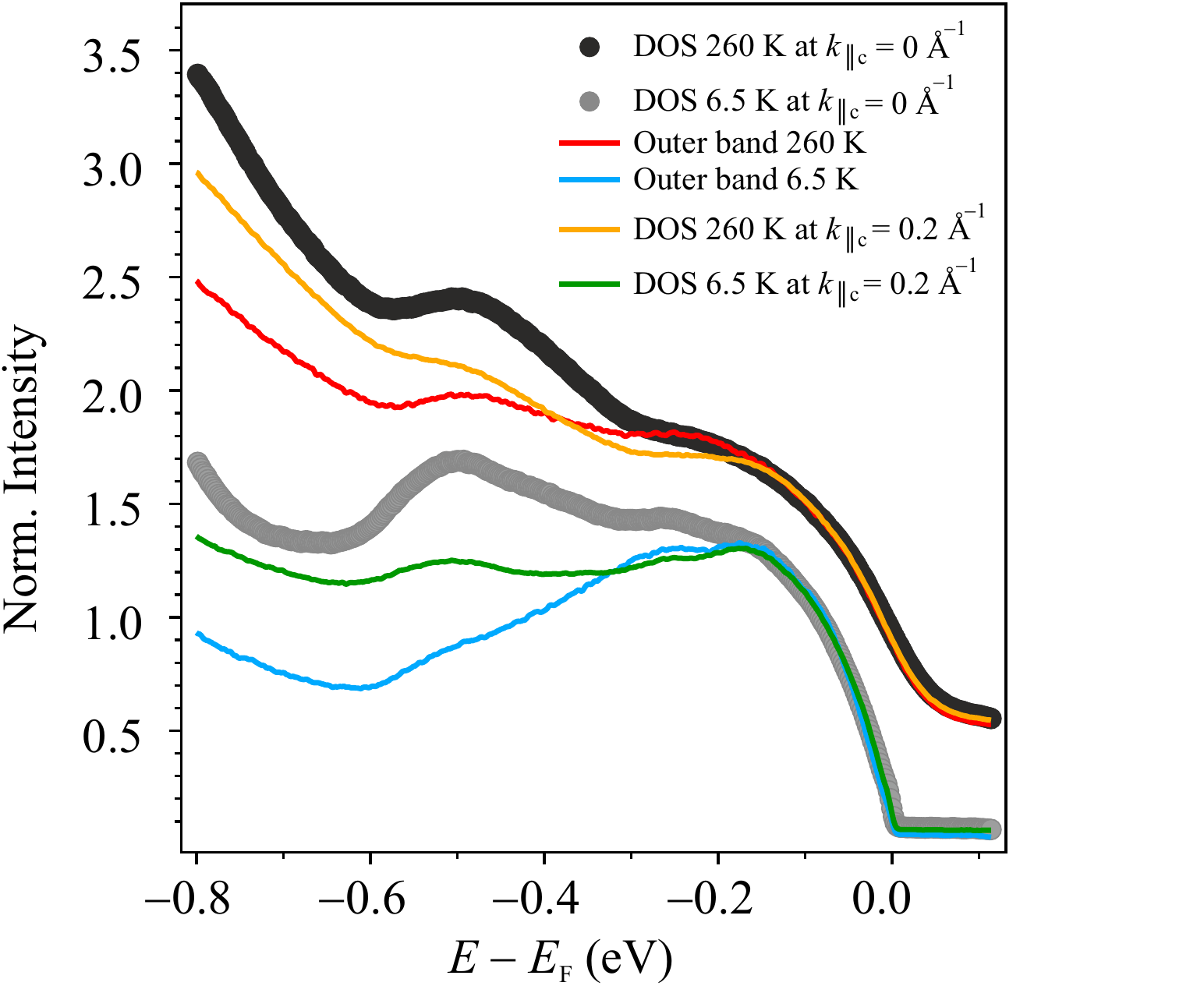}
\caption{\label{fig:DOS}
Integrated DOS as in Fig.~4b of the main text at 260~K and 6.5~K compared with the intensity in the outer band only (red and blue) in the momentum range $k_{\parallel \mathrm{b}}=-0.35$ to $-0.2~\mathrm{\AA}^{-1}$. The integrated DOS at $k_{\parallel \mathrm{c}}=0.2~\mathrm{\AA}^{-1}$ are also shown for comparison (orange and green). All curves are normalised at $-0.15$~eV and curves at 260~K are offset vertically for clarity.}
\end{figure}

The integrated DOS as in Fig.~4b of the main text at 260~K and 6.5~K is shown in Fig.~\ref{fig:DOS} and is compared with the intensity in the outer band only in the momentum range $k_{\parallel \mathrm{b}}=-0.35$ to $-0.2~\mathrm{\AA}^{-1}$. The integrated DOS at $k_{\parallel \mathrm{c}}=0.2~\mathrm{\AA}^{-1}$ is also shown for comparison. The shape of the DOS does not vary in the range of our analysis. From this we conclude that the analysis presented in the main text does not depend strongly on the $k_{\parallel \mathrm{b}}$ momentum range considered, nor on the $k_{\parallel \mathrm{c}}$ momentum.

\section*{\boldmath Tomonaga Luttinger liquid spectral function and Luttinger parameter $K_{\rho}$}

The microscopic description of a temperature-induced dimensional crossover requires minimally a quasi-1D Hamiltonian with a small transverse kinetic energy $t_{\perp}$. We are not aware of an analytical solution which would capture the evolution of the DOS in such a quasi-1D system as a function of the crossover parameter $T/t_{\perp}$. At $T\gg t_{\perp}$, we are in the 1D regime: the equivalent 1D Hamiltonian has only two parameters, an effective Fermi velocity and an effective Luttinger coefficient $K_{\rho}$ which both depend on the parameters of the original quasi-1D Hamiltonian, including $t_{\perp}$. At $T\ll t_{\perp}$, we expect an anisotropic Fermi liquid (FL) with a featureless DOS. Therefore, the DOS evolves from a power law at high $T$ to a constant at low $T$. In a Tomonaga-Luttinger liquid (TLL), a similar evolution is achieved by varying $K_{\rho}$ in the range $0<K_{\rho}\leqslant 1$. Our analysis assumes that the dimensional crossover in the quasi-1D model as a function of $T/t_{\perp}$ and/or $E/t_{\perp}$ can be mapped onto a TLL to FL transition in a strictly 1D model with varying $K_{\rho}$.

\begin{figure}[t]
\includegraphics[width=\columnwidth]{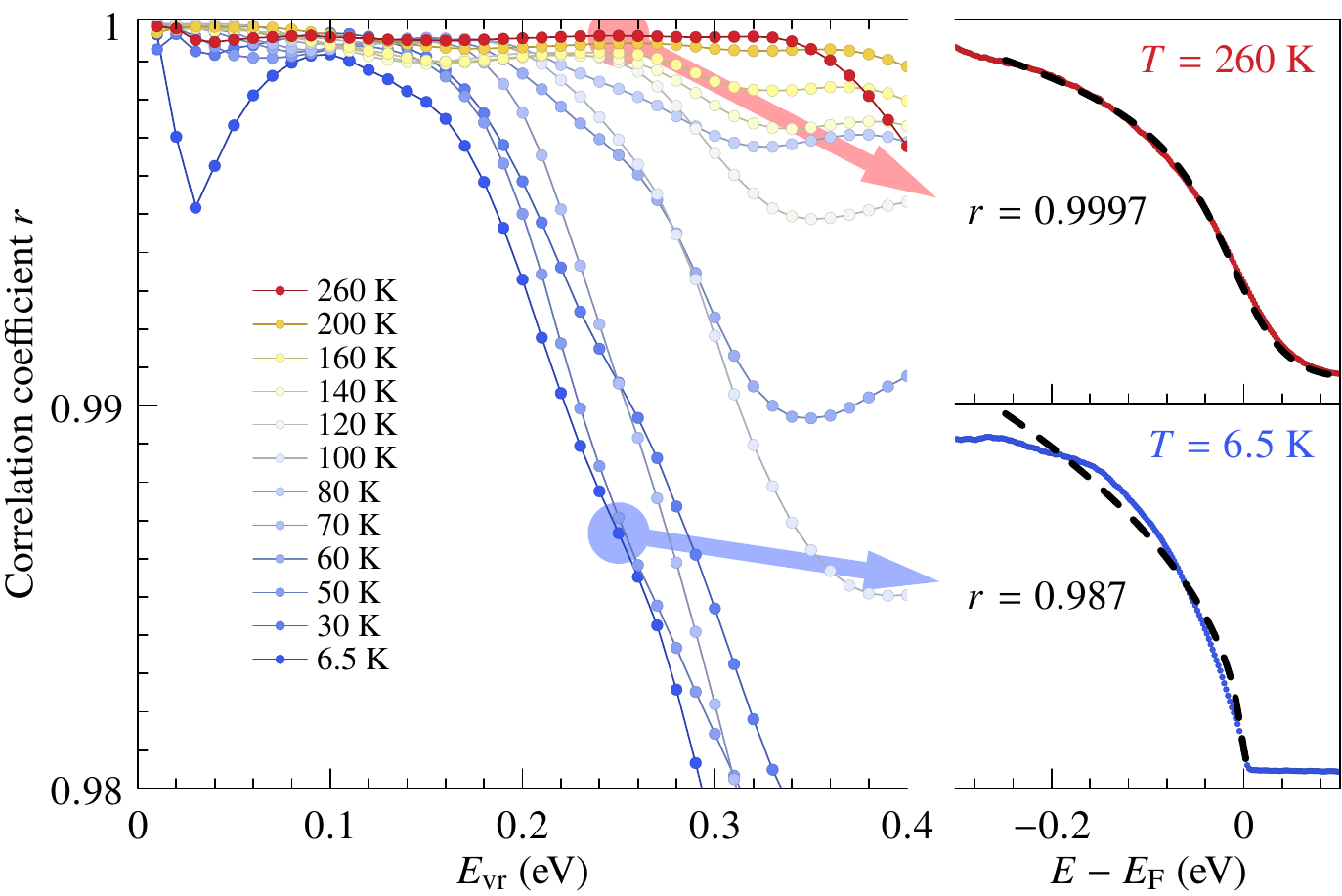}
\caption{\label{fig:correlation}
Correlation coefficient of the fit of Eq.~(\ref{eq:DOS}) to the momentum-integrated ARPES intensity in the energy range $[-E_{\mathrm{vr}},\min(E_{\mathrm{vr}},4k_{\mathrm{B}}T)]$. The resulting values of the exponent $\alpha$ for $T\geqslant100$~K are shown in Fig.~4c of the main text. Two examples of fits with high and low correlation are displayed on the right.
}
\end{figure}

An analytical expression is available for the temperature-dependent DOS of a TLL \cite{Schonhammer1993s}. We therefore analyze the integrated ARPES intensity, as was already done in Refs.~\onlinecite{Blumenstein2011s, Ohtsubo2015s}, using the expression
	\begin{equation}\label{eq:DOS}
		I(E,T)\propto T^{\alpha}\cosh\left(\frac{\varepsilon}{2}\right)
		\left|\Gamma\left(\frac{1+\alpha}{2}+i\frac{\varepsilon}{2\pi}\right)\right|^2
		f(\varepsilon)\ast g(E).
	\end{equation}
$\varepsilon=E/k_{\mathrm{B}}T$, $\Gamma$ is the Euler gamma function, and $f(\varepsilon)=(e^\varepsilon+1)^{-1}$ is the Fermi distribution. The symbol $\ast$ stands for a convolution with the instrumental resolution, represented by a Gaussian $g(E)$ of width 10~meV (FWHM). The exponent $\alpha$ describes the asymptotic zero-temperature DOS of the TLL behaving as $|E|^{\alpha}$. It is related to the microscopic parameters $K_{\rho}$ and $K_{\sigma}$ controlling the algebraic decay of correlation functions in the charge and spin sectors, respectively, by \cite{Giamarchi2003s}
	\begin{equation}\label{eq:alpha}
		\alpha=\frac{K_{\rho}^{\phantom1}+K_{\rho}^{-1}+
			K_{\sigma}^{\phantom1}+K_{\sigma}^{-1}}{4}-1.
	\end{equation}
For spin-rotation invariant systems like \NbSe{} we have $K_{\sigma}=1$, such that $\alpha=(K_{\rho}+K_{\rho}^{-1}-2)/4$ and we can deduce the value of $K_{\rho}$ from the fitted exponent $\alpha$:
	\begin{equation}\label{eq:Krho}
		K_{\rho}=1+2\alpha-2\sqrt{\alpha(\alpha+1)}.
	\end{equation}
In the pure 1D model (\ref{eq:DOS}), $\alpha$ relates to the strength of the microscopic interactions. Our interpretation is that the value of $\alpha$ obtained by fitting Eq.~(\ref{eq:DOS}) to the DOS of a quasi-1D system like \NbSe{} can indicate the evolution from a 1D regime where $\alpha>0$ to a 3D regime where $\alpha=0$, as the temperature and/or the energy is lowered.

We extract an optimized exponent $\alpha$ by least squares fitting in a variable energy range extending from $-E_{\mathrm{vr}}$ to $\min(E_{\mathrm{vr}},4k_{\mathrm{B}}T)$. The Pearson correlation coefficient is plotted in Fig.~\ref{fig:correlation} as a function of $E_{\mathrm{vr}}$ for all temperatures in our data set. As explained in the main text, we consider this fitting as meaningful only for temperatures higher than the highest CDW transition temperature (145~K) and energies higher than the dimensional crossover scale (110~meV). In this region the correlation coefficient is typically above 0.997. At low temperature, the fit worsens at large $E_{\mathrm{vr}}$ because of the CDW gaps in the DOS, while at low $E_{\mathrm{vr}}$---where the fit quality improves due to reduced number of data points---the resulting exponent increases, revealing an inconsistency with the 1D model (see main text). As a result the curves of $\alpha$ vs $E_{\mathrm{vr}}$ in Fig.~4c of the main text are not shown for $T<100$~K.

\end{document}